\documentclass{webofc}
\usepackage[varg]{txfonts}   % Web of Conferences font
\newcommand{\GeVc}{\ensuremath{\mbox{Ge\kern-0.1em V}\!/\!c}\xspace}
\newcommand{\GeV}{\ensuremath{\mbox{Ge\kern-0.1em V}}\xspace}

\newcommand{\AGeV}{\ensuremath{A\,\mbox{Ge\kern-0.1em V}}\xspace}
\newcommand{\AGeVc}{\ensuremath{A\,\mbox{Ge\kern-0.1em V}\!/\!c}\xspace}

\begin{document}

\setlength{\belowcaptionskip}{-5pt}
\title{NA61/SHINE overview}

\author{\firstname{Piotr} \lastname{Podlaski}\inst{1}\fnsep\thanks{\email{piotr.podlaski@cern.ch}}\\ \textit{for the NA61/SHINE Collaboration}
        % etc.
}

\institute{Faculty of Physics, University of Warsaw, Warsaw, Poland
          }
\abstract{%
  NA61/SHINE is a multipurpose fixed-target experiment at the CERN Super Proton Synchrotron. The main goals of the NA61/SHINE strong interaction program are to discover the critical point of strongly interacting matter as well as to study the properties of produced particles relevant for the study of the onset of deconfinement -- the transition between the state of hadronic matter and the quark-gluon plasma. An analysis of hadron production properties is performed in proton-proton, proton-nucleus and nucleus-nucleus interactions as a function of collision energy and size of the colliding nuclei to achieve these goals.
  
  The NA61/SHINE results from a strong interaction measurement program are presented. In particular, the latest results from different reactions Ar+Sc, Xe+La, and Pb+Pb on hadron spectra and fluctuations are discussed. Additionally, the recent measurements of kaon production in Ar+Sc collisions at $\sqrt{s_{NN}}=11.94~\GeV$ are mentioned. The NA61/SHINE results are compared with worldwide experiments and predictions of various models, like EPOS, PHSD, UrQMD, and others. 
}
\maketitle
\section{Introduction}
\label{sec:intro}
\begin{figure}[ht]
    \centering
    \includegraphics{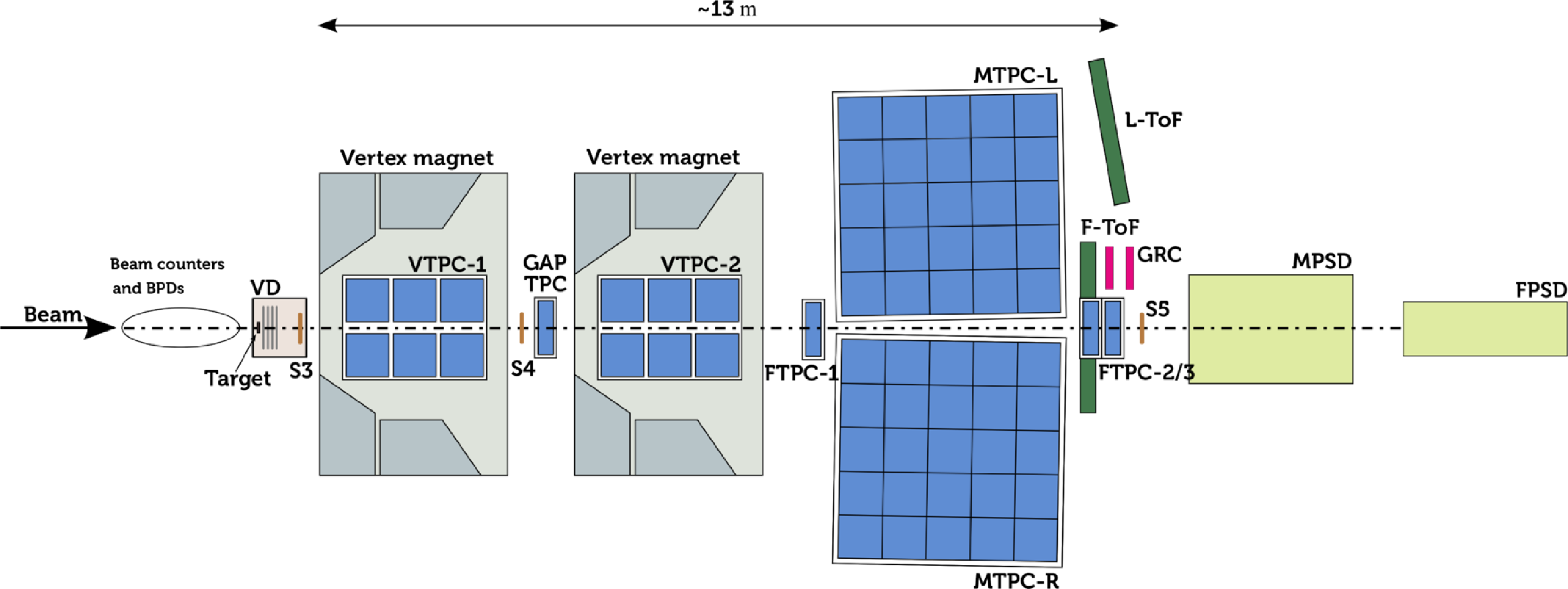}
    \caption{Schematic layout of  the upgraded NA61/SHINE detector system.}
    \label{fig:detector}
\end{figure}

NA61/SHINE is a large acceptance hadron spectrometer located in the CERN's North Area~\cite{fac_paper}. Eight large volume Time Projection Chambers (TPC), accompanied by the Time of Flight detectors (ToF) provide tracking and identification of produced particles. The Projectile Spectator Detector (PSD), a precise zero-degree hadron calorimeter, measures the energy of projectile spectators, which can be related to the centrality of the collision. The NA61/SHINE detector system was recently upgraded: the readout rate was increased beyond 1~kHz, a new Vertex Detector, as well as data acquisition and trigger systems were installed, upgrade of PSD was performed. The layout of the upgraded detector is shown in Fig. \ref{fig:detector}.\\
For strong interactions program NA61/SHINE performed a unique, two-dimensional scan in system size ($p$+$p$, $p$+Pb, Be+Be, Ar+Sc, Xe+La, Pb+Pb) and momentum of the beam (13$A$--150(8)\AGeVc) aiming on studies of the phase diagram of strongly interacting matter and search for the critical point.

\section{Search for the critical point}

\begin{figure}[hb]
    \centering
    \includegraphics[width=0.45\textwidth]{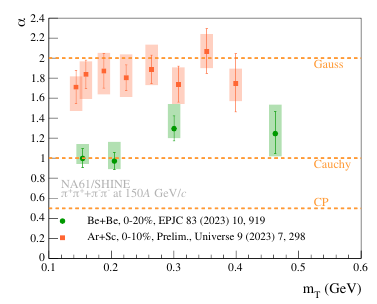}
    \includegraphics[width=0.45\textwidth]{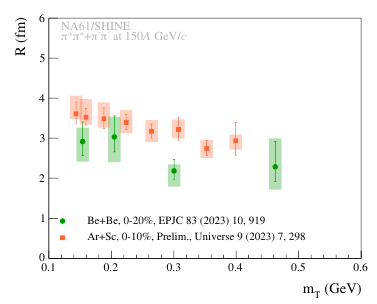}
    \caption{\textit{Left}: Lévy stability parameter $\alpha$ versus $m_T$ in 0–20\% central Be+Be and 0--10\% central Ar+Sc collisions at 150\AGeVc. Lines represent Gaussian shape with $\alpha=2$, Cauchy shape with $\alpha=1$, and vicinity of the CP at $\alpha\leq 0.5$. \textit{Right}: Lévy scale parameter R versus $m_T$ in 0--20\% central Be+Be and 0--10\% central Ar+Sc collisions at 150\AGeVc. }
    \label{fig:femto}
\end{figure}

The expected signal of a critical point (CP) is a non-monotonic dependence of various fluctuation/correlation measures in the NA61/SHINE energy-system size scan.

The nature of the quark-hadron transition can be studied via femtoscopy analysis as the investigation of the femtoscopic correlation functions in nucleus-nucleus reactions may reveal the space-time structure of the hadron production source. With the use of Lévy-type sources ($C(q)=1+\lambda e^{-|qR|^\alpha}$), we can describe the source parameters as a function of the average transverse mass of the pion pair \cite{femto_ref}.

The results of one-dimensional two-pion femtoscopic correlation measurements for identified pion pairs ($\pi^+\pi^++\pi^-\pi^-$) in Be+Be \cite{femto_bebe} and Ar+Sc \cite{femto_arsc} collisions at 150\AGeVc with 0--20\% and 0--10\% centrality are depicted in Fig. \ref{fig:femto}. The pion-producing source in the 150\AGeVc Be+Be collision appears compatible with the Lévy shape assumption, deviating significantly from Gaussian ($\alpha=2$) and closely resembling Cauchy ($\alpha=1$). In Ar+Sc at 150\AGeVc, $\alpha$ values range between 1.5 and 2.0, notably higher than in the case of Be+Be and closer to the Gaussian source shape. In both cases, obtained $\alpha$ values surpass the conjectured CP value ($\alpha\leq0.5$). The measured Lévy scale parameter $R$ demonstrates a decrease with increasing $m_T$ for both analyzed systems, which is attributed to the transverse flow of the evolving system.

\begin{figure}[t]
    \centering
    \includegraphics[width=0.48\textwidth, trim = 0 0 10 0, clip]{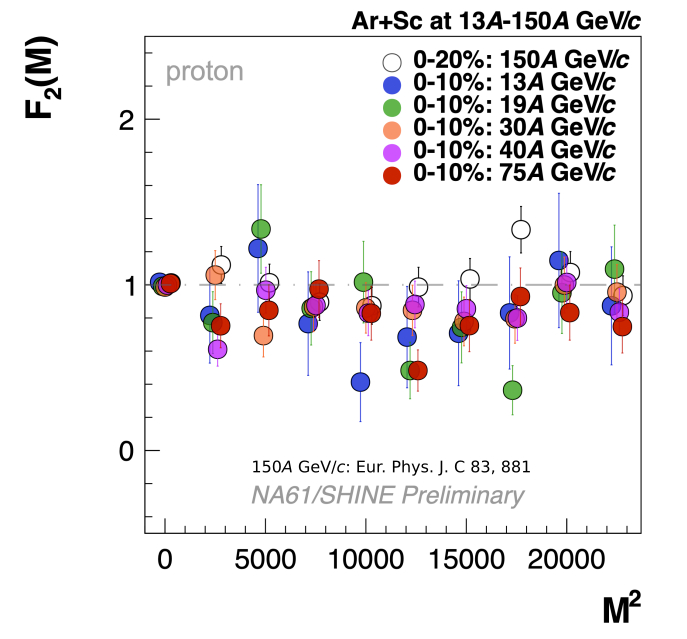}
    \includegraphics[width=0.45\textwidth, trim = 0 0 10 0, clip]{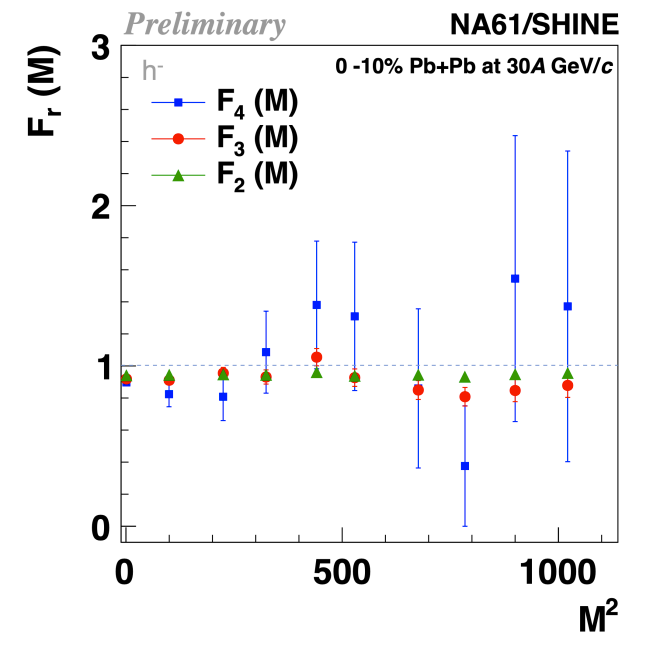}
    \caption{\textit{Left}: Results on $F_2(M)$ of protons in central Ar+Sc collisions at 13$A$--150\AGeVc. \textit{Right}: Scaled factorial moments of 2--4 order for negatively charged hadrons produced in the 10\% most central Pb+Pb collisions at 30\AGeVc}
    \label{fig:inter}
\end{figure}

Another tool for CP investigation is proton intermittency. In the proximity of the CP, local power-law fluctuations of the baryon density are expected. This can be explored by studying scaled factorial moments with the cell size or, equivalently, with the number of cells in space of protons at mid-rapidity \cite{Bialas:1985jb,Diakonos:2006zz}. NA61/SHINE measures scaled factorial moments of multiplicity distributions $F_r(M)$ using statistically independent points and cumulative variables. Results on $F_2(M)$ of mid-rapidity protons measured in central Ar+Sc collisions at 13$A$--150\AGeVc \cite{inter_high_mom,SR2023} are presented in the left panel of Fig. \ref{fig:inter}. Additionally, results on $F_r(M)$ ($r=2,3,4$) of negatively charged hadrons in the 10\% most central Pb+Pb collisions at 30\AGeVc \cite{haradan_pb} are shown in the right panel of Fig. \ref{fig:inter}. The power-law scaling exponent ($F_2(M)\sim M^{2\phi_2}$) intermittency index $\phi_2$ for a system freezing out at the QCD critical endpoint is expected to be $\phi_2 = 5/6$, assuming it belongs to the 3-D Ising universality class. Measured $F_2(M)$ of protons for Ar+Sc at 13$A$--150\AGeVc shows no indication of a power-law increase with bin size, which could indicate CP. This holds for $F_2(M)$ of negatively charged hadrons for Pb+Pb at 30\AGeVc as well.

\section{Studies of the onset of deconfinement}
\label{sec-ood}
According to the Statistical Model of The Early Stage (SMES) \cite{smes1,smes2} the first-order phase transition between Quark-Gluon Plasma (QGP) and hadron gas should be located in the CERN SPS energy range. 
\begin{figure}[ht]
\centering
\includegraphics[width=0.45\textwidth]{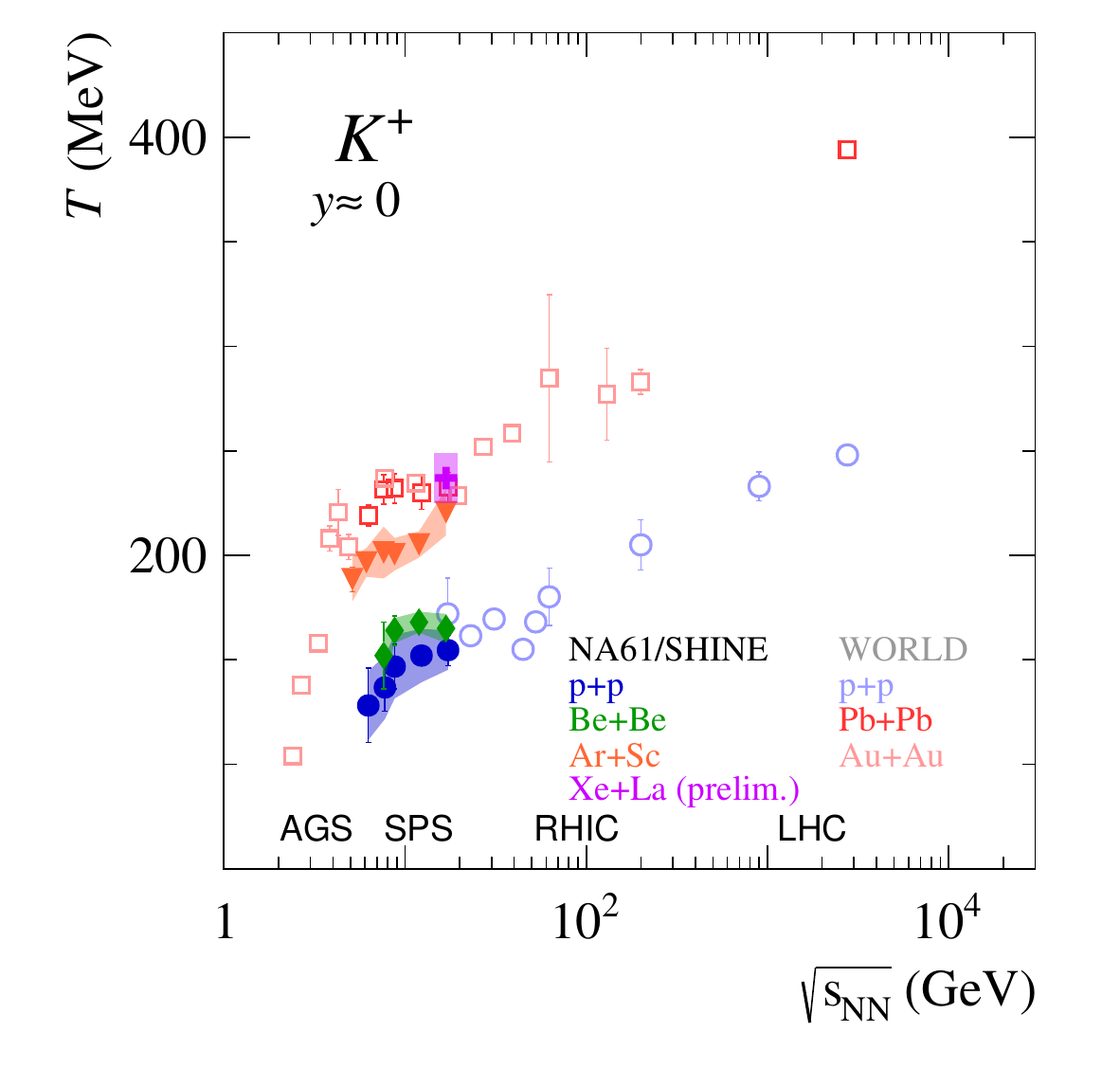}~~~~
\includegraphics[width=0.45\textwidth]{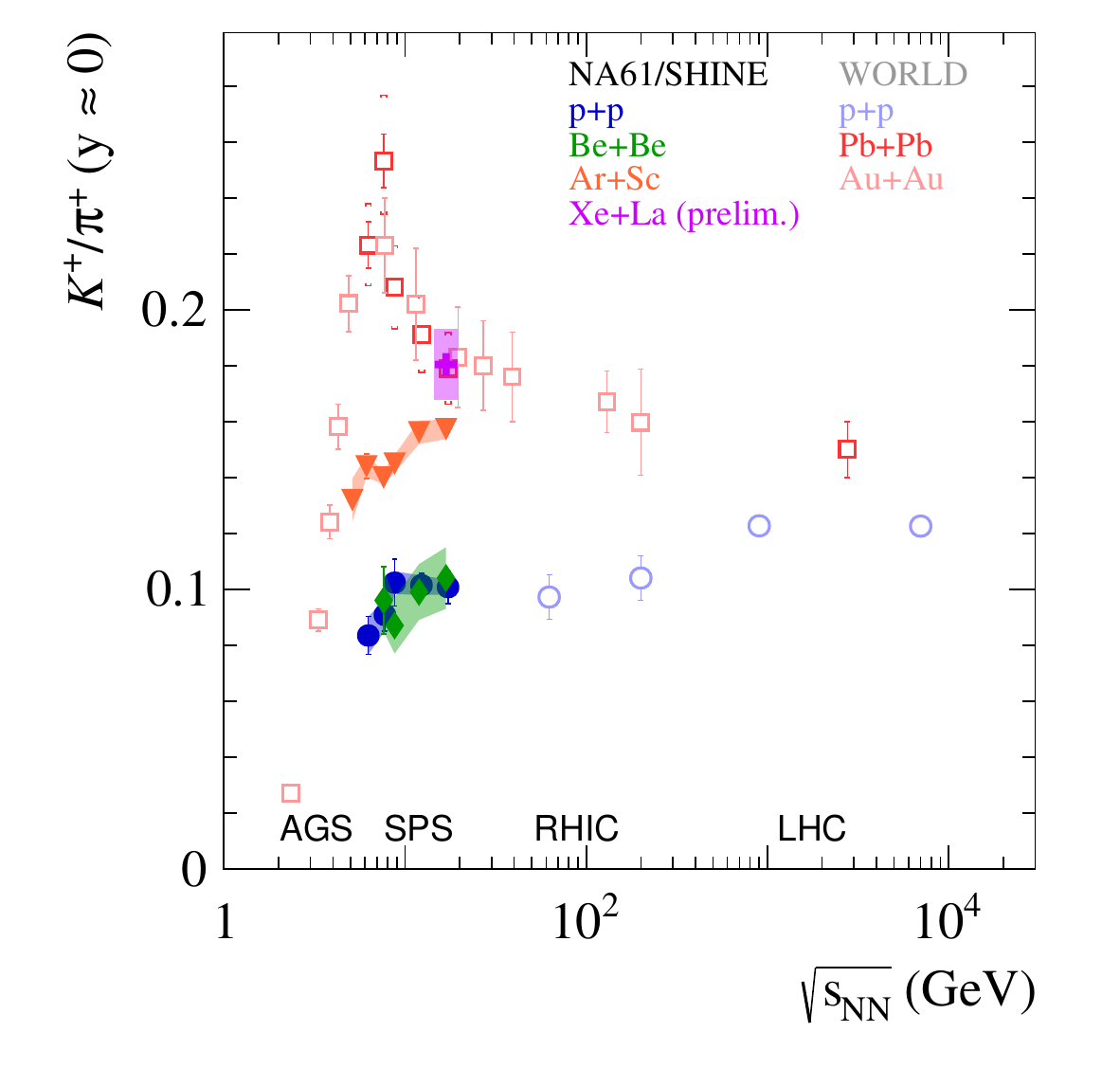}
\vspace{-0.45cm}

\caption{\textit{Left}: Energy dependence of the inverse slope parameter of $K^+$. \textit{Right}: Energy dependence of and of the $K^+/\pi^+$ particle yield ratio at mid-rapidity for $p$+$p$, Be+Be, Ar+Sc, Xe+La, and Pb+Pb collisions \cite{arsc_paper,sasha_poster}.}
\label{fig-step-horn}
\end{figure}
In the left panel of Fig. \ref{fig-step-horn}, the inverse slope parameter of positive kaon $m_T$ spectra at mid-rapidity is displayed for various colliding systems as a function of collision energy. According to SMES, the characteristic plateau observed in the energy dependence of the inverse slope parameter for heavy ion collisions (Pb+Pb, Au+Au) signifies the existence of the mixed phase of QGP and hadron gas. Notably, NA61/SHINE results for light and intermediate mass systems ($p$+$p$, Be+Be, Ar+Sc) exhibit a similar trend, with the plateau value growing with the system size. The inverse slope parameter $T$ of $K^+$ produced in Xe+La collisions at the top SPS collision energy aligns with the Pb+Pb result. The most prominent signature of the phase transition predicted within SMES, is a rapid, non-monotonic change of the $K^+/\pi^+$ ratio as a function of collision energy, the horn. The right panel of Fig. \ref{fig-step-horn} presents a compilation of results on positive kaon to positive pion multiplicity ratio at mid-rapidity. The horn structure was not observed in data on central Ar+Sc collisions. Moreover, Ar+Sc results on $K^+/\pi^+$ ratio at low SPS energies are located between light ($p$+$p$, Be+Be) and heavy (Pb+Pb, Au+Au) systems. As collision energy increases, the $K^+/\pi^+$ ratio in Ar+Sc approaches the values for heavy systems. Remarkably, at the top SPS energy, the $K^+/\pi^+$ ratio in Xe+La mirrors that of Pb+Pb.

\begin{figure}[h]
\centering
\includegraphics[width=0.45\textwidth]{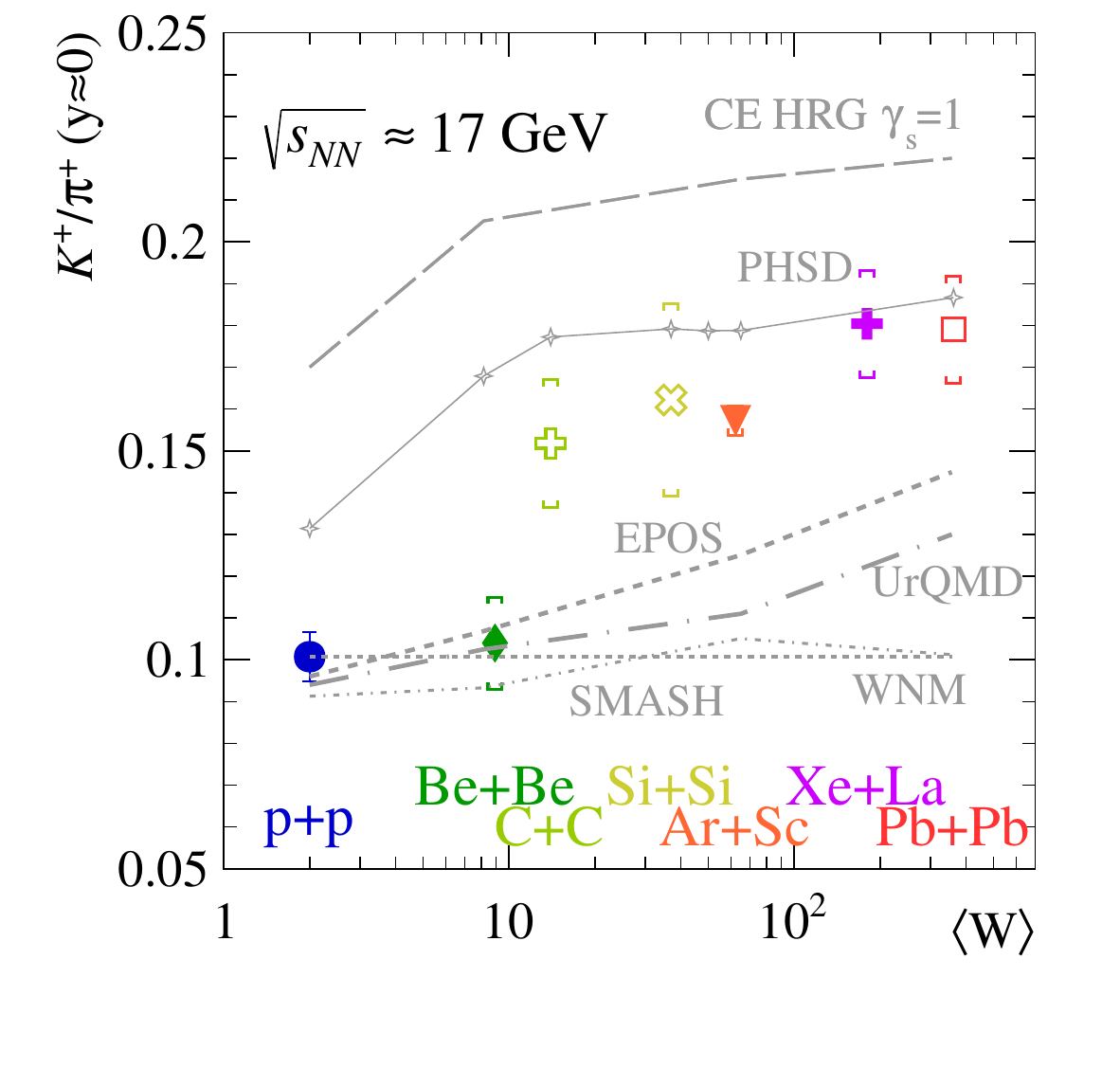}
\includegraphics[width=0.45\textwidth]{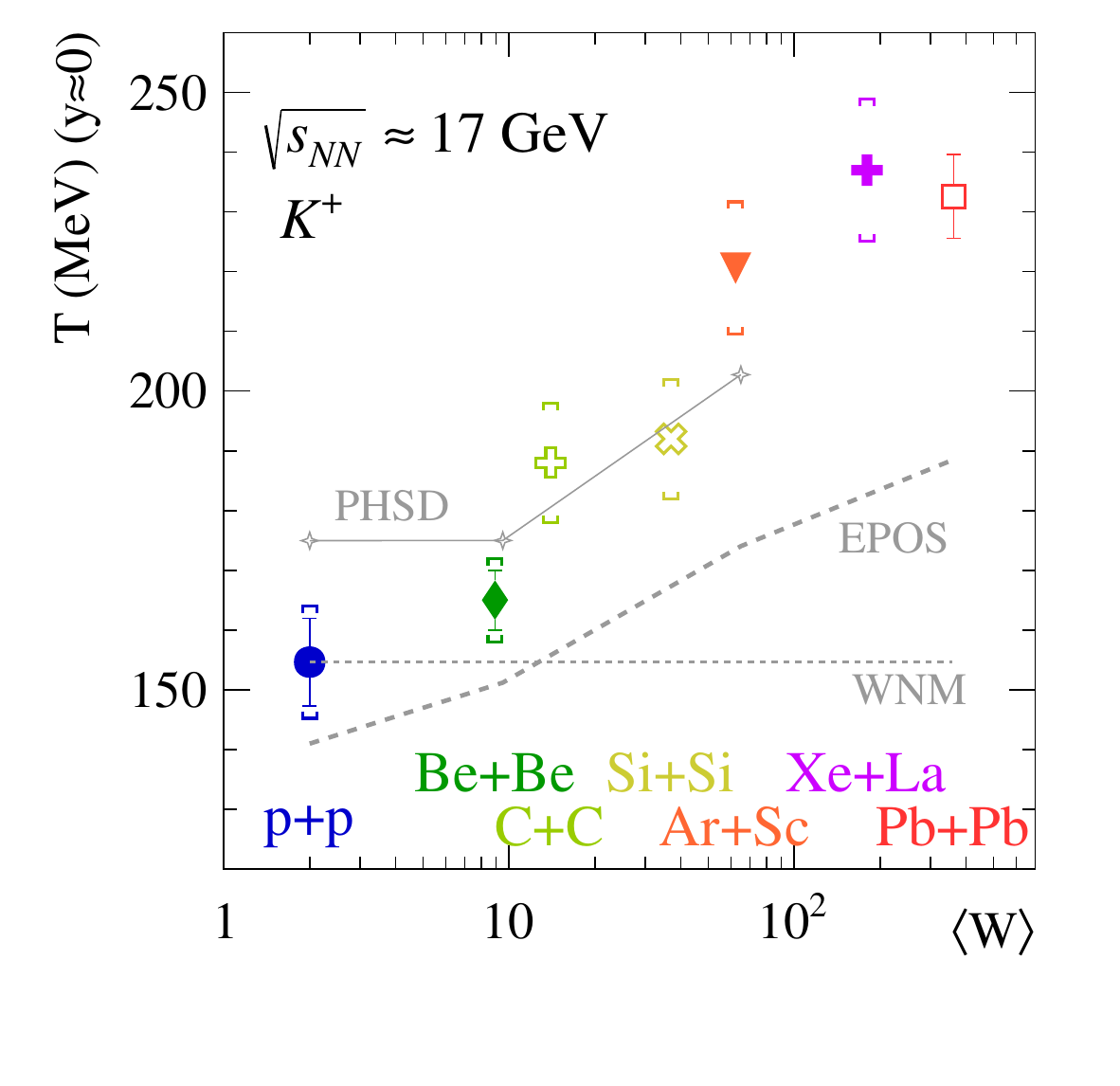}
\vspace{-0.4cm}
\caption{\textit{Left}: System size dependence of the $K^+/\pi^+$ ratio at mid-rapidity measured at 150(8)\AGeVc compared with models. \textit{Right}: System size dependence of the inverse slope parameter $T$ of $K^+$ at the same collision energy \cite{arsc_paper,sasha_poster}.}
\label{fig-syst-size}
\end{figure}

Figure \ref{fig-syst-size} presents $K^+/\pi^+$ multiplicity ratio and inverse slope parameter of the $K^+$ transverse mass spectra as a function of the system size for the top SPS energy (150\AGeVc beam momentum). The mean number of wounded nucleons $\langle W \rangle$ quantifies system size of a collision. Dynamical models (EPOS, UrQMD, SMASH) successfully describe the $K^+/\pi^+$ ratio for light systems ($p$+$p$ and Be+Be) but fail for heavier ones (Ar+Sc, Xe+La, Pb+Pb). On the other hand, PHSD, the model with phase transition, reproduces the data for heavy systems but overestimates the $K^+/\pi^+$ ratio for lighter ones. The Hadron Resonance Gas model (HRG) tends to overestimate the data. Both considered quantities ($K^+/\pi^+$ ratio and inverse slope parameter $T$) show qualitatively similar system size dependence, which cannot be fully described by any of the statistical or dynamical models used for comparison.

\section{Large isospin symmetry violation in kaon production?}

%The NA61/SHINE collaboration conducted measurements on $K^0_S$ production in 0–10\% central Ar+Sc collisions at 75\AGeVc \cite{kzero_paper}. The resulting $K^0_S$ rapidity and transverse momentum spectra are illustrated in Fig. \ref{fig:kzero_results}. The mean multiplicity of produced $K^0_S$ mesons was determined by integrating the fitted rapidity function, yielding $\langle K^0_S\rangle=6.25 \pm 0.09 (\text{stat}) \pm 0.73 (\text{sys})$.

\begin{figure}[ht]
    \centering
    \includegraphics[width=\textwidth]{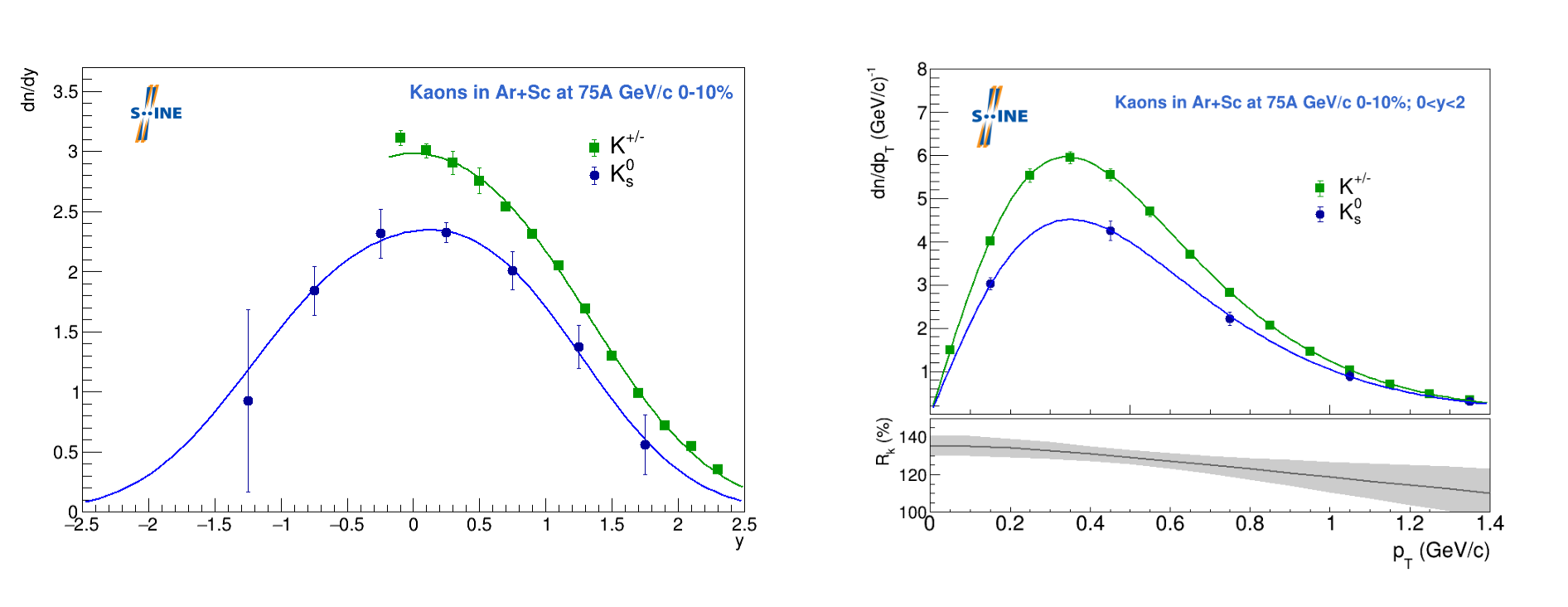}
    \caption{\textit{Left}: Comparison of rapidity spectrum of neutral ($K^0_S$) with the average spectrum of charged ($K^+$ and $K^-$) mesons in 0–10\% central Ar+Sc collisions at 75\AGeVc. The total uncertainties are plotted and calculated as the square root of the sum of squared statistical and systematic uncertainties $\left(\sqrt{(\text{stat})^2+(\text{sys})^2}\right)$. For charged kaons \cite{arsc_paper}, the total uncertainties were calculated separately for positively charged and negatively charged kaons and then propagated. \textit{Right}: same as left but for transverse momentum spectra.}
    \label{fig:kzero_results}
\end{figure}

\begin{figure}[hb]
    \centering
    \includegraphics{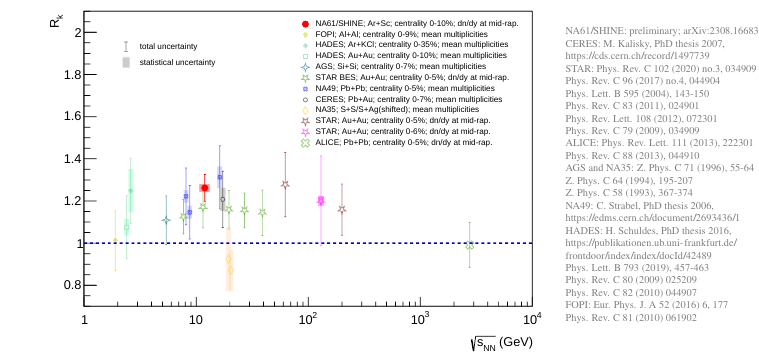}
    \caption{Compilation of the available data on the ratio of charged to neutral kaons as a function of collision energy. The measurement from NA61/SHINE is shown as a red dot. The world data needed to obtain $R_k$ values come from the references given in the right panel of the figure.}
    \label{fig:chrged_neutral_kaons}
\end{figure}

Strong interactions preserve approximately isospin ($I$) and its third component ($I_z$), which, among others, for collisions of $N=Z$ nuclei ($N$ -- number of neutrons, $Z$ -- number of protons) corresponds to equivalence in the production of new pairs of $u-\overline{u}$ and $d-\overline{d}$ quarks~\cite{ref_Pal}. Following Smushkevich rule, for all particles involved in isospin-conserving reactions, all members of isospin multiplets are produced in equal numbers if and only if the initial population is uniform~\cite{ref_Smushkevich, ref_Wohl, MacFarlane:1965wp}. Thus, for an electric-to-baryon charge ratio ($Q/B$; $Q \equiv Z$) equal to 1/2 ($I$ = $I_z$ = 0) and in the case of exact isospin symmetry we expect the following relations between kaon multiplicities:
$K^+(u\overline{s})  = K^{0}(d\overline{s})$ and $K^-(\overline{u}s) = \overline{K}^{0}(\overline{d}s)$. By summing up the equations one obtains: $K^+ + K^- = K^{0} + \overline{K}^{0}$. The $K^{0}$ and $\overline{K}^{0}$ mesons are not directly measured in detectors since the physical neutral states are the \ensuremath{K^{0}_{S}} and \ensuremath{K^{0}_{L}}. Neglecting a very small effect of the CP violation, the production of \ensuremath{K^{0}_{S}} should be given by: $K^{0}_{S} = \frac{K^{0} + \overline{K}^{0}}{2}$. Therefore, we expected the relation between multiplicities: $K^{0}_{S} = \frac{ K^+ + K^-}{2}$.

The left panel of Fig.~\ref{fig:kzero_results} shows the comparison of rapidity spectrum of neutral ($K^0_S$) \cite{wojtek_proc} with the average spectrum of charged ($K^{+}$ and $K^{-}$)~\cite{arsc_paper} mesons ($K^{+/-} = \frac{K^+ + K^-}{2}$). A similar plot but for transverse momentum spectra is presented in the right panel of Fig.~\ref{fig:kzero_results}. Additionally, for transverse momentum spectra, the $R_k$ ratio is plotted, where $R_k = \frac{K^{+/-}}{K^0_{S}}$. A significant difference between $K^{+/-}$ and $K^0_S$ yields is observed for both rapidity and transverse momentum spectra.

%In Figure \ref{fig:chrged_neutral_kaons}, a compilation of data depicting the ratio of charged to neutral kaons is presented as a function of collision energy. The data reveals a systematic excess in the production of charged kaons within the showcased nucleus-nucleus collisions. Despite the large uncertainties associated with individual data points in the world dataset, they align with and substantiate the findings observed by NA61/SHINE. The experimental result was compared with the prediction of HRG in \cite{kzero_th}. The prediction of HRG takes into account a set of known effects that violate isospin symmetry, or preserve it but still can lead to a deviation of $R_k$ from unity. Nevertheless, the predicted deviation from the expected value of $R_k=1$ is significantly smaller than for the experimental data. Thus, the presented result is to be considered as evidence for effects that go beyond the ones predicted by the HRG model.

Figure \ref{fig:chrged_neutral_kaons} displays a compilation of data illustrating the ratio of charged to neutral kaons as a function of collision energy. The data reveals a systematic excess in the production of charged kaons within the showcased nucleus-nucleus collisions. Despite the considerable uncertainties associated with individual data points in the global dataset, they align with and support the observations made by NA61/SHINE. The experimental findings were compared with the predictions of models in Ref. \cite{kzero_th}. The models consider various known effects that either violate isospin symmetry or, while preserving it, may still lead to a deviation of $R_k$ from unity. However, the predicted deviation from the expected value of $R_k=1$ is notably smaller in the models than in the experimental data. This suggests that the presented results may indicate the presence of effects that go beyond those anticipated by the models.

\section*{Acknowledgments}
This work was partially supported by the Polish Minister of Science and Higher Education (contract~No.~2021/WK/10)

\bibliography{bibliography}

\end{document}